# PRESOLAR GRAINS FROM NOVAE

SACHIKO AMARI, XIA GAO[1], LARRY R. NITTLER[2] AND ERNST ZINNER

*Laboratory for Space Sciences and the Physics Department, Washington University, St. Louis, MO 63130-4899, USA; sa@howdy.wustl.edu, ekz@howdy.wustl.edu*

JORDI JOSÉ[3] AND MARGARITA HERNANZ

*Institut d'Estudis Espacials de Catalunya (IEEC/CSIC), E-08034 Barcelona, Spain; jjose@ieec.fcr.es, hernanz@ieec.fcr.es*

and

ROY S. LEWIS

*Enrico Fermi Institute, University of Chicago, Chicago IL 60637-1433, USA; r-lewis@uchicago.edu*

## ABSTRACT

We report the discovery of five SiC grains and one graphite grain isolated from the Murchison carbonaceous meteorite whose major-element isotopic compositions indicate an origin in nova explosions. The grains are characterized by low $^{12}$C/$^{13}$C (4-9) and $^{14}$N/$^{15}$N (5-20) ratios, large excesses in $^{30}$Si ($^{30}$Si/$^{28}$Si ratios range to 2.1 times solar) and high $^{26}$Al/$^{27}$Al ratios. These isotopic signatures are theoretically predicted for the ejecta from ONe novae and cannot be matched by any other stellar sources. Previous studies of presolar grains from primitive meteorites have shown that the vast majority formed in red giant outflows and supernova ejecta. Although a classical nova origin was suggested for a few presolar graphite grains on the basis of $^{22}$Ne enrichments, this identification is somewhat ambiguous since it is based only on one trace element. Our present study presents the first evidence for nova grains on the basis of major element isotopic compositions of single grains. We also present the results of nucleosynthetic calculations of classical nova models and compare the predicted isotopic ratios with those of the grains. The comparison points toward massive ONe novae if the ejecta are mixed with material of close-to-solar composition.

---

[1]Present address: The Scripps Research Institute, La Jolla, CA 92037, USA; xiagao@scripps.edu

[2]Present address: Laboratory for Extraterrestrial Physics, Code 691, NASA Goddard Spaceflight Center, Greenbelt MD 20771, USA; nittler@lepvax.gsfc.nasa.gov

[3]Departament de Física i Enginyeria Nuclear, Universitat Politècnica de Catalunya, E-08800 Barcelona, Spain; jordi.jose@upc.es

## 1. INTRODUCTION

Primitive meteorites contain presolar grains that condensed in stellar atmospheres and ejecta and survived interstellar travel and the formation of the solar system (Anders & Zinner 1993; Zinner 1998). Their study provides information on stellar evolution, nucleosynthesis and the physical and chemical conditions in stellar outflows and explosions (Bernatowicz & Zinner 1997). Most grain types identified so far are carbonaceous (diamond, silicon carbide, graphite, titanium carbide); oxide grains (corundum, spinel, titanium oxide, hibonite) and silicon nitride are much more rare (Zinner 1998; Nittler & Alexander 1999; Choi, Wasserburg, & Huss 1999). Asymptotic Giant Branch (AGB) stars, before but mostly during the thermally pulsing phase, have been identified as stellar sources of the majority of SiC and oxide grains. Supernovae are apparently the sources of a small fraction of SiC grains and of $Si_3N_4$ grains (Nittler et al. 1995; Hoppe et al. 2000), of low-density graphite grains (Travaglio et al. 1999), and of a few corundum grains (Nittler et al. 1998; Choi et al. 1998), and nanodiamonds carry a supernova signature in their Xe and Te isotopes (Lewis et al. 1987; Richter, Ott, & Begemann 1998).

Wolf-Rayet stars and classical novae have also been invoked as sources of presolar grains, but evidence has been much less direct. Although classical novae are only a minor contributor of nuclei and dust to the Galaxy, they are known to produce dust in their ejecta and have long been implicated as a possible source of some isotopic anomalies observed in meteorites, namely excesses in $^{26}$Mg (from radioactive decay of $^{26}$Al) and $^{22}$Ne (from $^{22}$Na decay). Searches for the carriers of these and other isotopic anomalies ultimately did lead to the identification of presolar circumstellar grains (stardust) in meteorites. However, so far the argument for the presence of grains with a nova origin rests mainly on the Ne isotopic composition of graphite grain aggregates (Amari, Lewis, & Anders 1995a): the high $^{22}$Ne/$^{20}$Ne ratio of the Ne-E(L) carried by these grains indicates the decay of short-lived $^{22}$Na produced in novae (Clayton 1975).

Ion microprobe isotopic analysis of individual SiC grains has revealed the existence of different populations that are distinguished by their C, N and Si isotopic ratios (Zinner 1998; Hoppe & Ott 1997). Figures 1 and 2 show the distributions of these ratios for the different SiC grain types. It should be emphasized that the numbers of the grains plotted in these two figures do not correspond to their abundances in primitive meteorites because rare grain types such as A+B, X, Y and Z grains have been preferentially located by isotopic imaging searches in the ion microprobe (Nittler 1996; Nittler et al. 1997). The abundances of the different types are indicated in Fig. 1. Mainstream, Y and Z grains are believed to come from AGB stars with a range of metallicities (Zinner et al. 2000), X grains from supernovae. J-type carbon stars and CH stars have been proposed as sources of SiC grains of type A and B (Lodders & Fegley Jr. 1998; Amari et al. 2000). In this paper, we report on five SiC grains and one graphite grain whose isotopic signatures differ from those of all other grains and strongly suggest that they formed in novae.

## 2. ISOTOPIC RATIOS OF NOVA GRAIN CANDIDATES AND THEIR STELLAR SOURCE

### 2.1. Isotopic Ratios

The isotopic properties of the grains are listed in Table 1. They have low $^{12}C/^{13}C$ ratios, extremely low $^{14}N/^{15}N$ ratios, huge $^{30}Si$ excesses and close-to-solar $^{29}Si/^{28}Si$ ratios (Figs. 1 and 2). We included SiC grain KJC112 (Hoppe et al. 1996) because, although its Si isotopic ratios have not been determined, its C and N ratios are very similar to those of the other grains in the Table. We also included graphite grain KFC1a-551 with its large $^{30}Si$ excess and low $^{12}C/^{13}C$ ratio (Amari, Zinner, & Lewis 1995b). Its $^{14}N/^{15}N$ ratio is solar but the majority of graphite grains have solar N isotopic ratios in spite of a large range in $^{12}C/^{13}C$ ratios and it has been argued that graphite grains suffered isotopic equilibration (Hoppe et al. 1995).

Because astronomical measurements of isotopic ratios in stars, especially for N, Al, and Si, the elements considered here, are extremely limited or non-existent, we have to rely on theoretical models of nucleosynthesis in different types of stars for identifying the most likely stellar source of presolar grains with known isotopic ratios of certain elements. It is well known that present nucleosynthetic models of essentially all stars are far from perfect and fraught with uncertainties and we will not dwell on this question. However, comparison of isotopic grain data with theoretical models has been fruitful in the past (e.g., Gallino, Busso, & Lugaro 1997) and in most cases it is all we can do. It should actually be realized that the grain data provide important constraints on models and thus contribute to our understanding of stellar nucleosynthesis.

### 2.2. AGB Stars

It is difficult if not impossible to explain the isotopic compositions of the grains in Table 1 if they originated from AGB stars or from supernovae. AGB stars are not predicted to have both low $^{12}$C/$^{13}$C and $^{14}$N/$^{15}$N ratios. After the 1$^{st}$ and 2$^{nd}$ dredge-up episodes, $^{12}$C/$^{13}$C and $^{14}$N/$^{15}$N ratios are expected to be 20-30 and 500-850, respectively (El Eid 1994; Boothroyd & Sackmann 1999). Low-mass stars ($<2M_{\odot}$) are believed to experience extra mixing, which results in lower $^{12}$C/$^{13}$C but in even higher $^{14}$N/$^{15}$N ratios (Charbonnel 1995; Wasserburg, Boothroyd, & Sackmann 1995). During the 3$^{rd}$ dredge-up the envelope becomes $^{12}$C-rich, enabling condensation of SiC and graphite, but naturally $^{12}$C/$^{13}$C ratios increase and there is little change in the N isotopic ratio.

### 2.3. Type II Supernovae

Type II supernovae are also unlikely sources. A minor population of presolar grains, SiC of type X, low-density graphite, and $Si_3N_4$, are believed to have formed in supernovae (Amari & Zinner 1997), but their isotopic signatures, high $^{12}$C/$^{13}$C and $^{26}$Al/$^{27}$Al, low $^{14}$N/$^{15}$N ratios and $^{28}$Si excesses, are completely different from those of the present grains. The isotopic signatures of the grains of this study, especially the combination of low $^{12}$C/$^{13}$C and low $^{14}$N/$^{15}$N ratios, are neither predicted for individual supernova zones (Woosley & Weaver 1995), nor for mixtures (Travaglio et al. 1999).

### 2.4. Wolf-Rayet Stars

Although there exist a couple of presolar grains that might have a Wolf-Rayet star origin (Nittler et al. 1997; Amari, Zinner, & Lewis 1999), such a stellar source is highly unlikely for the grains of this study because high $^{12}$C/$^{13}$C and low $^{14}$N/$^{15}$N ratios, as well as excesses of both $^{29}$Si and $^{30}$Si are expected for the WC phase of Wolf-Rayet stars (see, e.g., Arnould et al. 1997).

### 2.5. Classical Novae

In contrast, the basic isotopic signatures — low $^{12}$C/$^{13}$C and $^{14}$N/$^{15}$N ratios (Fig. 3), and high $^{26}$Al/$^{27}$Al ratios (Fig. 4) — are qualitatively consistent with the signatures predicted for novae. Several computations of nucleosynthesis in classical nova outbursts have been published during the last thirty years. However, only few extended analyses involving a representative number of isotopes and nuclear reactions, directly linked to hydrodynamic models, have been reported.

Parametrized and semi-analytic nova models (i.e., Hillebrandt & Thielemann 1982; Wiescher et al. 1985; Weiss & Truran 1990; Nofar, Shaviv & Starrfield 1991; Boffin et al. 1993; Coc et al. 1995; Wanajo, Hashimoto & Nomoto 1999), while successful in exploring a wide parameter space, usually ignore important ingredients in nova modeling, which deeply influence the accompanying nucleosynthesis. Examples include the early accretion phase, the progression of the outburst, the extent and subsequent recession of convection throughout the envelope - by far the most critical aspect, ignored in most simplified models - and the expansion and ejection stages. Hereafter, we will focus only on one-dimensional hydrodynamic calculations of nova outbursts (see Starrfield 1998, José, Coc & Hernanz 1999, Kovetz & Prialnik 1997 and references therein). We stress that such hydrodynamic models constitute the present state of the art in nova modeling and are able to account for some observational properties of classical nova outbursts. However, they suffer from important uncertainties (mainly the mixing process between the accreted material and the outermost shells of the white dwarf). While a detailed discussion of such models and their success exceeds the scope of this paper, the reader can find extended treatments addressed to these particular aspects elsewhere (see, for instance, Starrfield 1989, or José & Hernanz 1998, and references therein). So far, multidimensional treatments, which probably will solve some of the long-standing problems in nova modeling (see Shankar, Arnett & Fryxell 1992; Shankar & Arnett 1994; Glasner & Livne 1995; Glasner, Livne & Truran 1997; Kercek, Hillebrandt & Truran 1998, 1999) have not been successful in reproducing the gross properties of nova outbursts, the reason being a limitation in spatial and temporal resolution imposed by the capabilities of present-day computers.

Astronomical observations indicate that classical novae fall into two categories (e.g., Starrfield, Gehrz, & Truran 1997; Gehrz et al. 1998). CO novae are explosions on carbon and oxygen-rich white dwarf (WD) stars, typically of mass $<1.1M_\odot$. Neon (or ONe) novae arise from WDs rich in oxygen and neon as the result of the burning of carbon ($M>\sim1.1M_\odot$). Nucleosynthesis in classical novae occurs via proton capture reactions at high temperatures ($T_{max}\leq3.5\times10^8$K). In spite of the uncertainties associated with existing nova models, all 1D hydrodynamic calculations converge onto essentially the same predictions concerning the C, N and Al isotopic ratios, i.e., low $^{12}C/^{13}C$ and high $^{26}Al/^{27}Al$ ratios, whereas a wide dispersion of $^{14}N/^{15}N$ ratios is obtained if both CO and ONe models are considered (Kovetz & Prialnik 1997; Starrfield et al. 1997; Starrfield et al. 1998; José, Hernanz, & Coc 1997; José & Hernanz 1998; José, Coc, & Hernanz 1999; José 2000). The corresponding isotopic ratios observed in the grains qualitatively agree with theoretical predictions for both CO and ONe novae.

The Si isotopic signature can potentially give us more specific information as to the nature of the parent stars of the grains (Fig. 5). Models of CO novae predict close-to-solar or lower-than-solar Si ratios (Starrfield et al. 1997; José & Hernanz 1998) and thus would exclude CO novae, which do not produce any $^{30}$Si excesses but only $^{29}$Si deficits, as potential sources of the grains. In contrast, predicted Si isotopic ratios for ONe novae vary for different hydrodynamic models (José & Hernanz 1998; Starrfield et al. 1998). Nuclear processes in ONe novae involve isotopes of heavier elements than those in CO novae, reflecting both the fact that ONe novae achieve higher temperatures and that they contain heavier seed nuclei such as $^{20}$Ne, $^{23}$Na and $^{24,25}$Mg. The general trend can be summarized as follows: as the WD mass increases, both $^{29}$Si/$^{28}$Si and $^{30}$Si/$^{28}$Si increase, resulting in Si with huge enrichments of $^{30}$Si and with $^{29}$Si/$^{28}$Si ratios close to solar in the high WD-mass models. According to the models, $M_{WD}$ should be at least 1.25$M_\odot$ to account for the $^{30}$Si excesses observed in the grains.

### 2.6. Problems

We have already discussed the general uncertainties associated with nova models and will not reiterate this topic. A comparison of the results of these models with the isotopic characteristics in the grains indicate that they condensed in the ejecta of ONe novae. Such an origin is not without its difficulties, however. First, the observed isotopic compositions are less extreme than the predicted values. This is shown for the C, N, Al and Si isotopic ratios in Figs. 3, 4 and 5, where we plot the nova grain data together with results of nucleosynthetic models of CO and ONe classical novae (Kovetz & Prialnik 1997; Starrfield et al. 1997; Starrfield et al. 1998; José & Hernanz 1998; José 2000). To achieve the ratios measured in the grains requires mixing of material synthesized in the nova outburst with unprocessed, isotopically close-to-solar material. One possibility is that this material came from the binary companion star, which is usually assumed to be on the main sequence. Another possibility is that the outermost layers of the envelope were ejected shortly after the onset of the explosion and that they contained less processed material.

In Figs. 3-5 we plot mixing lines between the predictions of two ONe models, one for a 1.25$M_\odot$ WD mass, the other for 1.35$M_\odot$, and the solar composition and indicate the mixing fraction of solar material. Because the mass fractions of C, N, Al, and Si in the ONe nova products are much higher than in material of solar composition (for Si $\sim$46 - 112 times as much as in material of solar composition for the ONe models by José & Hernanz 1998), more than 95% of isotopically close-to-solar material has to be mixed with model nova ejecta to achieve the observed grain compositions. The two model compositions, both for single ejecta shells, were chosen for demonstration purposes and do not imply that the grains necessarily obtained their compositions from a mix of a single shell with solar material. It is much more likely that several shells have to be mixed with material of solar composition.

From Fig. 3 it can be seen that the J00 models with 1.15$M_\odot$ and 1.25$M_\odot$ WD masses cover ranges so that appropriate mixing of these compositions and solar can account for the C and N isotopic ratios of the grains. In contrast, the J00 models with 1.35$M_\odot$ WD masses and the Starrfield et al. (1998) models (all with a WD mass of 1.25$M_\odot$) yield too low N ratios for a possible fit to the grain data. The $^{26}$Al/$^{27}$Al ratios predicted by ONe models with different WD masses cover similar ranges (Fig. 4) and at least the ratio of grain KJGM4C-311-6 can be explained by appropriate mixing. The Si isotopic ratios of at best grains AF15bC-126-3, KJGM4C-100-3, and KJGM4C-311-6 can be explained by mixing of J00 models with 1.25$M_\odot$ WD masses if we expand "solar compositions" to the Si isotopic compositions of mainstream SiC grains (Fig. 2), however, to explain the Si ratios of grain AF15bB-429-3 and graphite KFC1a-551 we need to mix 1.35$M_\odot$ J00 models or STWS models with solar. It is thus clear that no single model or mixing of different shell models of a given WD mass can consistently explain all (C, N, Al, Si) the isotopic compositions of a given grain.

A second problem is that theoretical nova models predict a higher abundance of oxygen than carbon in both CO and ONe novae, yet it is commonly assumed that C>O has to be satisfied in order for C-rich dust such as SiC or graphite to condense. Furthermore, spectroscopic observations on CO and ONe novae (e.g., Table 1 by Starrfield et al. 1998 and Table 4 by Wanajo et al. 1999 and references therein) have shown that nova ejecta are O-rich. However, C-rich dust has been observed around ONe novae, even those that also show evidence for silicate dust (see, e.g., review papers by Gehrz, Truran, & Williams 1993 and by Starrfield et al. 1997). These observations and our data thus suggest that either nova ejecta are clumpy, with both C-rich and O-rich regions (Gehrz et al. 1998), or that dust formation in novae does not occur under equilibrium conditions (Scott 2000). Clayton, Liu, & Dalgarno (1999) have pointed out that in supernova explosions the CO molecule, which normally controls condensation chemistry, is destroyed by radiation, freeing C atoms to possibly form carbonaceous dust even in the presence of larger amounts of oxygen. Perhaps a similar process occurs in the hard radiation field of novae.

Third, dust is commonly observed to condense in CO, but less frequently in ONe novae (Gehrz et al. 1986; Snijders et al. 1987; Woodward et al. 1992; Starrfield et al. 1997; Gehrz 1988; Mason et al. 1996). Furthermore, it is estimated that only 20-30% of novae are ONe novae (see e.g., Gehrz et al. 1993, 1998). However, comparison of the Si isotopic signatures of the putative nova grains with present-day models requires the nuclear processing to occur in ONe novae. While the astronomical data do not preclude some dust forming in ONe novae, one might expect to find more presolar grains from CO novae than from ONe, contrary to our results.

## 3. FUTURE DIRECTIONS

It would be highly desirable to locate more nova grain candidates and to perform more detailed isotopic measurements on them. Table 1 shows that not all relevant isotopic measurements could be made on the grains found to date. Unfortunately, such grains are exceedingly rare, in agreement with estimates how much novae contribute to the interstellar dust inventory (e.g., see paper by Alexander 1997). Furthermore, they seem to be smaller than SiC grains from AGB stars and supernovae. We are optimistic that more sensitive instrumentation such as the Cameca IMS 1270 and NanoSIMS ion microprobes (Stadermann, Walker, & Zinner 1999) will make it possible to perform isotopic measurements of many elements on the same grain, allowing, for example, the confirmation of high $^{26}$Al/$^{27}$Al ratios in nova grains.

It is also worthwhile to search for other isotopic signatures. Models predict large $^{33}$S excesses in ONe novae. So far, no S isotopic measurements have been made on presolar SiC grains but equilibrium condensation calculations indicate that various trace elements might be incorporated into SiC as sulfides (Lodders & Fegley Jr. 1995). Titanium would also provide a highly diagnostic isotopic signature. Since H burning does not affect such heavy elements, Ti isotopic ratios in nova grains are predicted to show at most the isotopic composition of material from the companion star, i.e. isotopic effects much smaller than the large $^{30}$Si excesses of nova grains.

Finally, oxide grains from novae would be clearly identified by enormous $^{17}$O and somewhat smaller $^{18}$O excesses. Corundum ($Al_2O_3$) grains are expected to exhibit large inferred $^{26}$Al/$^{27}$Al ratios and Al-Mg spinel is expected to carry pronounced excesses in $^{25}$Mg and $^{26}$Mg (Starrfield et al. 1997; Starrfield et al. 1998; José & Hernanz 1998), the latter being due to both nucleosynthetic production and decay of $^{26}$Al. Presolar oxide grains are rare and to date only grains from red giant and AGB stars (e.g., Nittler et al. 1997) and a few from supernovae (Nittler et al. 1998; Choi et al. 1998) have been located, making the possibility of finding oxides from novae a low-probability proposition. On the other hand, we are confident in our ability to identify more SiC grains with a nova origin and to study their isotopic composition in much more detail than has been possible so far.

We thank Roberto Gallino for general discussions on nucleosynthesis, James W. Truran for discussions on various aspects of nova explosions and suggestions on mixing with material of solar composition, and Alain Coc for some suggestions regarding nuclear uncertainties, which affect sulfur production in novae. This work has been supported by NASA grants NAG-8336 (S.A. and E.Z.) and NAG5-4297 (R.S.L.), by CICYT-P.N.I.E. grant ESP98-1348 and DGICYT grants PB98-1183-C02 and PB98-1183-C03 (J.J. and M.H).

Table 1. CANDIDATES OF GRAINS WITH A NOVA ORIGIN

| Grain | $^{12}$C/$^{13}$C $\pm 1\sigma$ | $^{14}$N/$^{15}$N $\pm 1\sigma$ | $\delta^{29}$Si/$^{28}$Si$^{*}$ $\pm 1\sigma$ | $\delta^{30}$Si/$^{28}$Si$^{*}$ $\pm 1\sigma$ | $^{26}$Al/$^{27}$Al |
|---|---|---|---|---|---|
| AF15bB-429-3[a] | 9.4±0.2 | | 28±30 | 1118±44 | |
| AF15bC-126-3[a] | 6.8±0.2 | 5.22±0.11 | -105±17 | 237±20 | |
| KJGM4C-100-3 | 5.1±0.1 | 19.7±0.3 | 55±5 | 119±6 | 0.0114 |
| KJGM4C-311-6 | 8.4±0.1 | 13.7±0.1 | -4±5 | 149±6 | >0.08 |
| KJC112[b] | 4.0±0.2 | 6.7±0.3 | | | |
| KFC1a-551[c] | 8.5±0.1 | 273±8 | 84±54 | 761±72 | |
| ONe nova models | 0.3∼3 | 0.1∼10 | -800∼1800 | -800∼15000 | 0.07∼0.7 |

$^{*}\delta^{i}$Si/$^{28}$Si $\equiv$ [($^{i}$Si/$^{28}$Si)$_{meas}$/($^{i}$Si/$^{28}$Si)$_{solar}$ - 1] $\times$ 1000

[a]Gao & Nittler (1997)

[b]Hoppe et al. (1996)

[c]Amari et al. (1995b)

Fig. 1.— Nitrogen and C isotopic ratios of individual presolar SiC grains. Different types of grains are distinguished by their C, N, Al, and Si isotopic ratios. The numbers in the legend give the abundances of the various grain types in primitive meteorites. The numbers of grains of different types plotted in figures 1 and 2 do not represent their natural abundances because rare grains have been selectively located by isotopic imaging in the ion microprobe (Nittler 1996). Also plotted is a graphite grain that, together with four SiC grains, is likely to have a nova origin. The dotted lines depict the solar isotopic ratios of 272 (N) and 89 (C).

Fig. 2.— Silicon isotopic ratios of presolar SiC grains and graphite grain KFC1a-551, a nova grain candidate. Ratios are plotted as $\delta$-values, deviations from the solar system ratios of $^{29}$Si/$^{28}$Si = 0.05063 and $^{30}$Si/$^{28}$Si = 0.03347 in permil (‰). The solid line with a slope of 1.31 represents the correlation line of the mainstream SiC grains (Lugaro et al. 1999). Grains of type Y and Z and nova grain candidates have $^{30}$Si excesses relative to this line but nova grains are distinguished by their low $^{12}$C/$^{13}$C and $^{14}$N/$^{15}$N ratios (Fig. 1) and high $^{26}$Al/$^{27}$Al ratios (Table 1).

Fig. 3.— Nitrogen and C isotopic ratios of four SiC and one graphite nova grain candidates are plotted together with those theoretically predicted for the ejecta from CO and ONe novae. The predicted ratios by Kovetz & Prialnik (1997) are indicated as KP97, Starrfield et al. (1997) as SGT97, José & Hernanz (1998) as JH98, Starrfield et al. (1998) as STWS98, and José (2000) as J00. Theoretical predictions vary widely, depending on parameters such as WD mass, WD temperature, mass transfer rate and the location of a given zone within the ejecta. In particular, CO nova ejecta can have large excesses but also large depletion of $^{15}$N relative to the solar composition. The grain data (because of wide-spread nitrogen isotopic equilibration in graphite the $^{14}$N/$^{15}$N ratio of the one graphite grain has to be considered an upper limit) fall between the solar system isotopic composition and the compositions of many ONe models. This indicates mixing between the products of nucleosynthesis in the nova explosion and isotopically close-to-solar material. In this figure as well as Figs. 4 and 5 the solid line represents the result of mixing between the predicted composition of an individual shell (i.e., not the total nova ejecta) of a ONe nova model with W$_{WD}$=1.25$M_\odot$ and the dotted line the result of mixing with a shell of a ONe nova model with W$_{WD}$=1.35$M_\odot$ (José 2000). The numbers along the lines show mass fractions of solar material in the mix. In order to match the grain data, more than 95% of the mix must have had close-to-solar composition. Also indicated in the figure are the ranges of the J00 single shell models for different WD masses and the range of the ONe models by Starrfield et al. (1998). Only a mixture of different shells of the J00 models with 1.15$M_\odot$ and 1.25$M_\odot$ WD masses with solar material can give a fit to the grain data.

Fig. 4.— Aluminum and C isotopic ratios of two nova grain candidates are plotted together with theoretical predictions for the ejecta from CO and ONe novae. The symbols in this figure as well as the mixing lines are the same as in Fig. 3. For one grain we obtained only a lower limit for the

$^{26}$Al/$^{27}$Al ratio, because we could not eliminate the Al signal from a nearby Al-rich oxide grain. The predicted ratios range from 0.005 to 1. Again, mixing between the nova ejecta and material with close-to-solar isotopic composition is necessary to account for the isotopic ratios of the grains. Also this figure indicates that solar material should comprise more than 95% of the mix.

Fig. 5.— Silicon isotopic ratios of nova grain candidates are plotted together with theoretical predictions for the ejecta from CO and ONe novae. Symbols and mixing lines are the same as in Fig. 3. Nucleosynthesis in CO novae produces $^{29}$Si deficits but no significant $^{30}$Si excesses. In contrast, mixing of ejecta material from ONe novae with a WD mass of at least $1.25 M_{\odot}$ with material of solar-system isotopic composition can, in principle, explain the Si isotopic ratios of the grains. However, no internal consistency with the C and N isotopic compositions is obtained (see text).

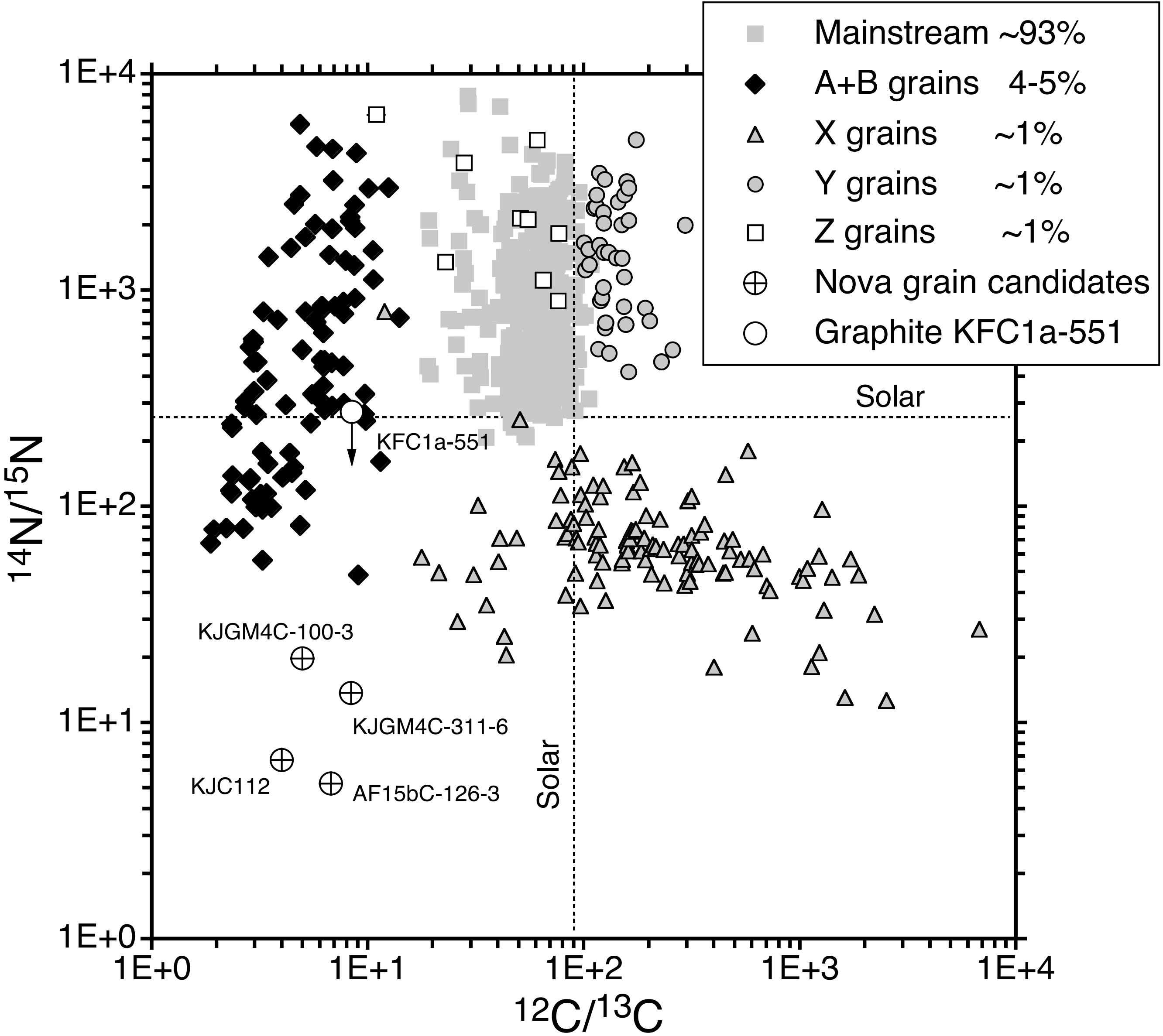

Figure 1

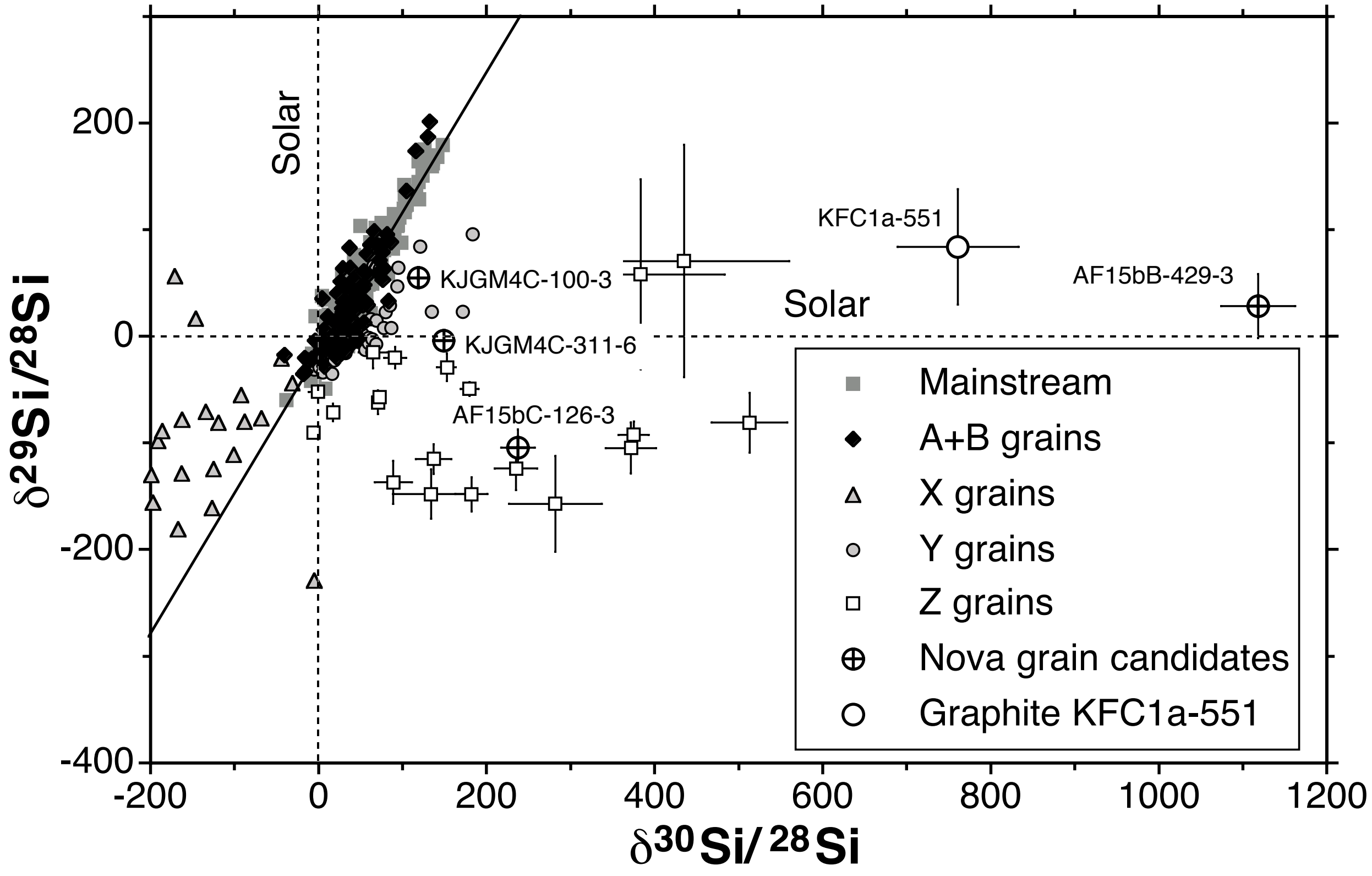

Figure 2

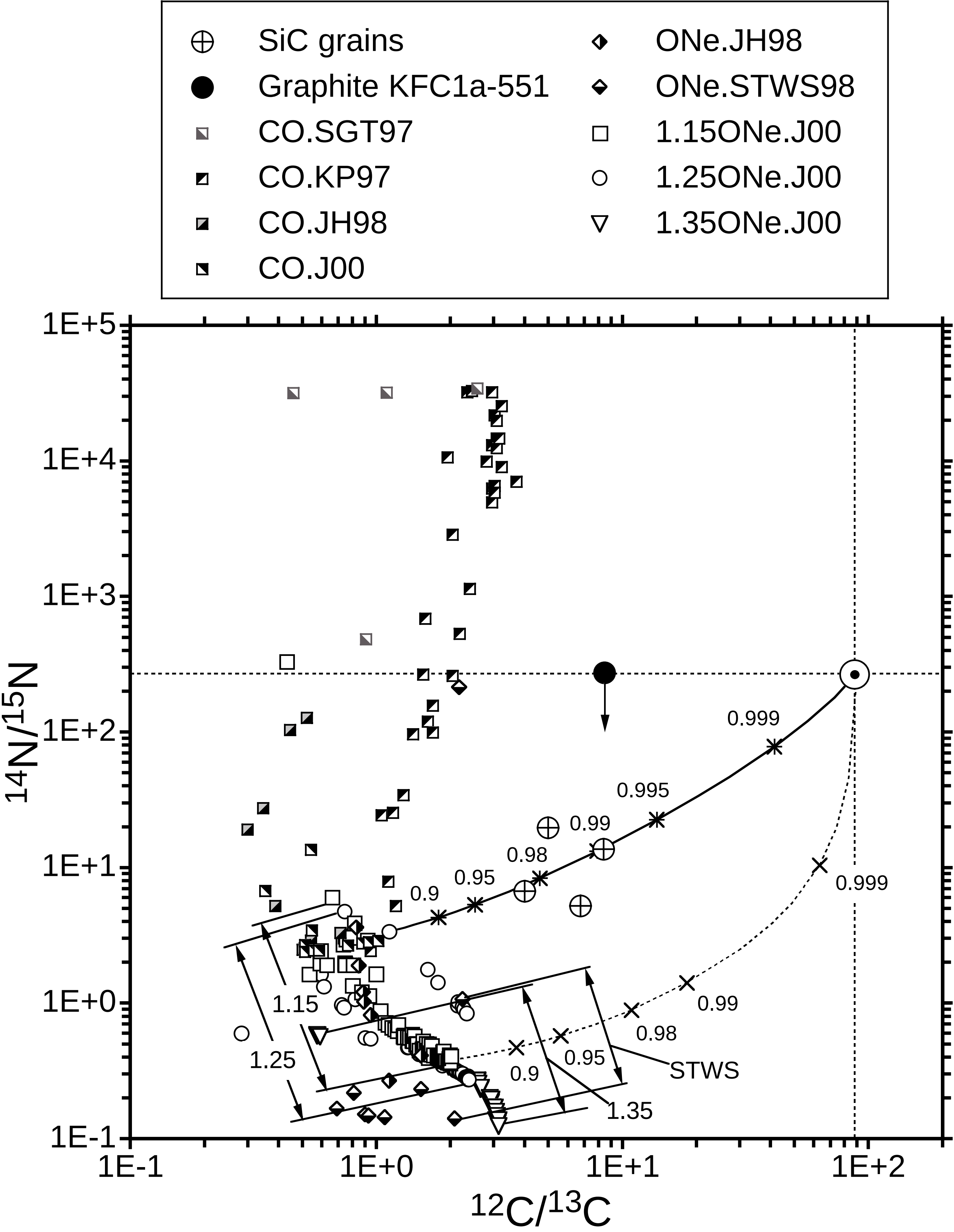

Figure 3

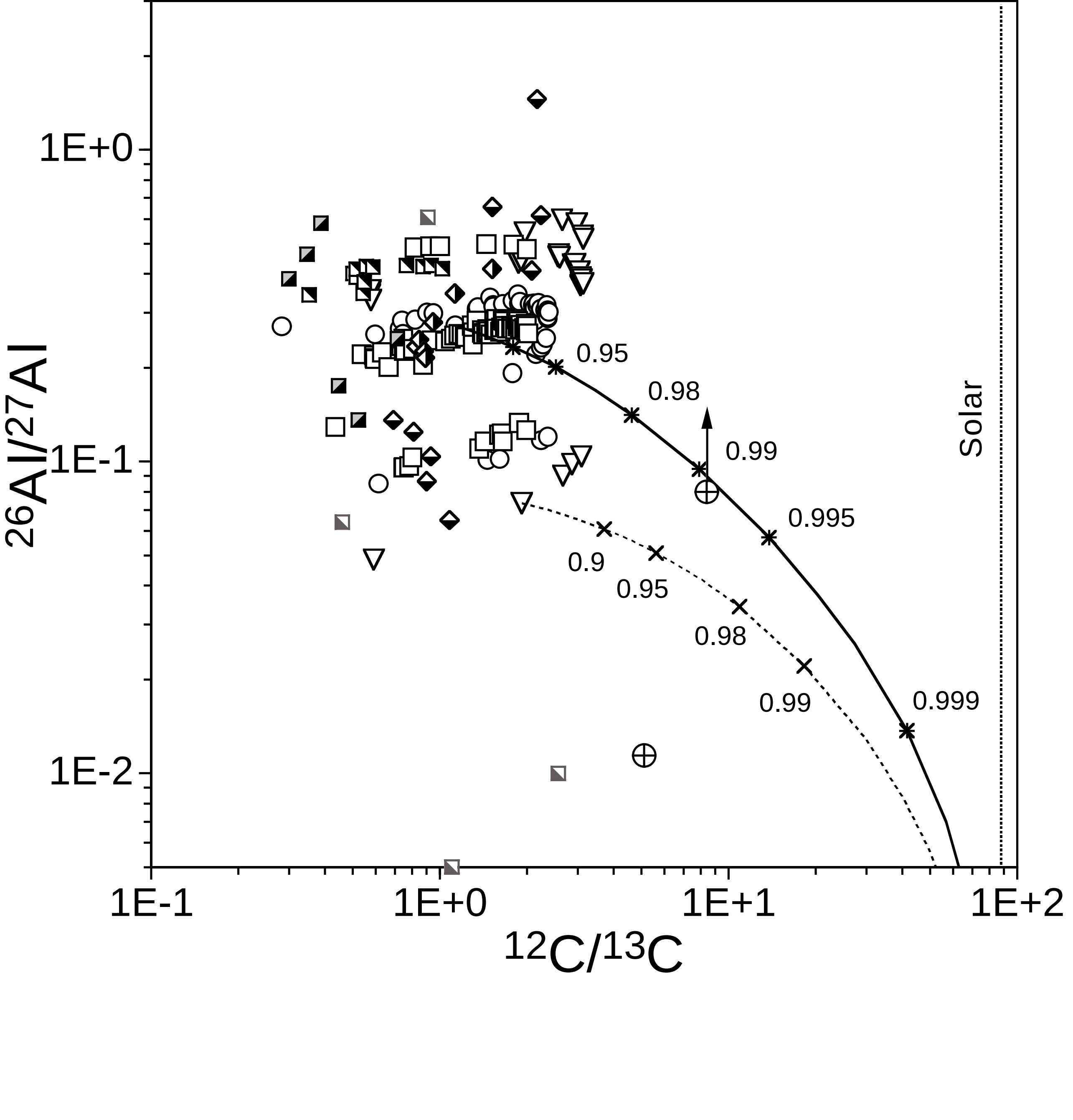

Figure 4

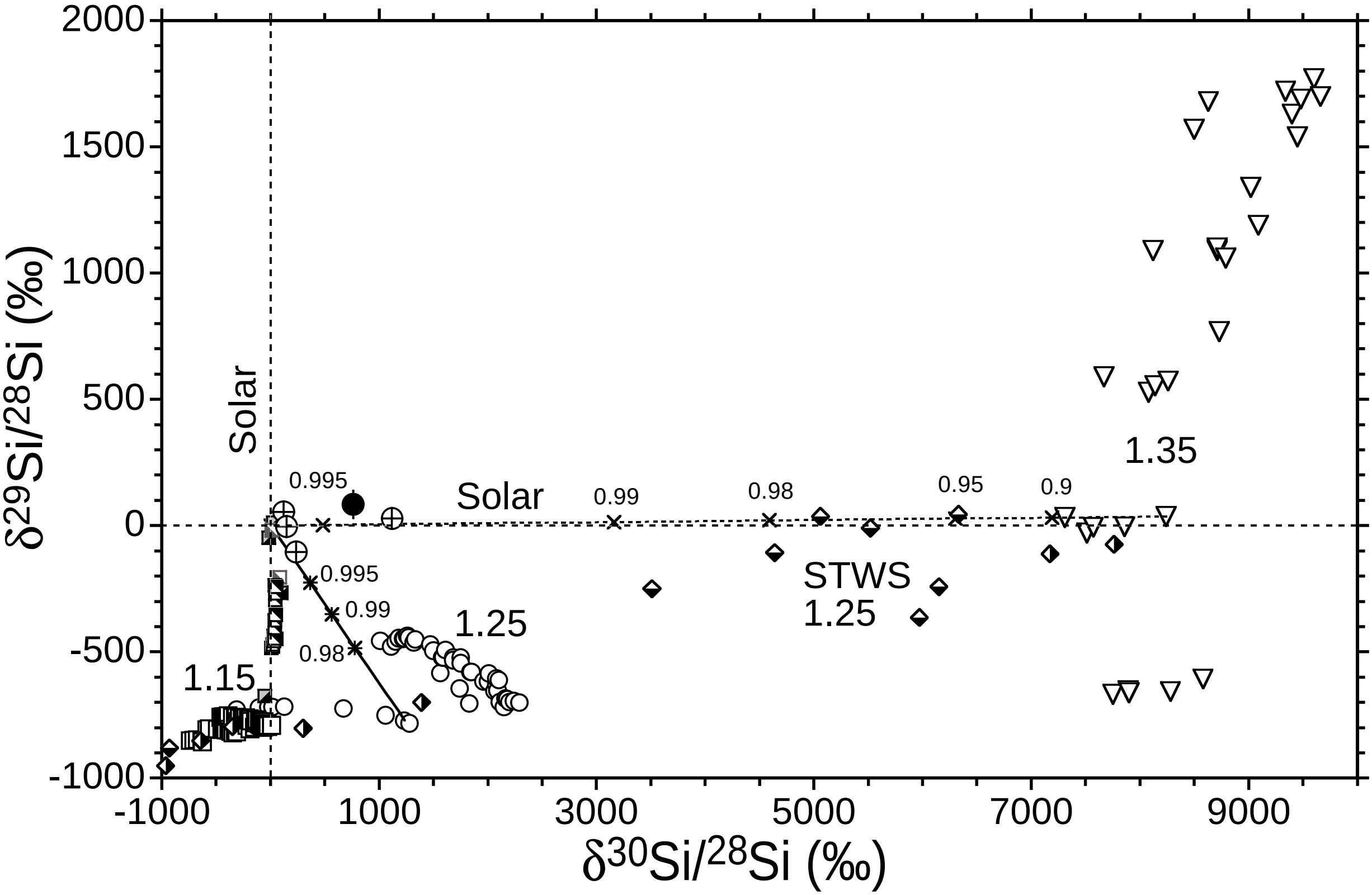

Figure 5